\documentclass{article}
\usepackage{arxiv}
\usepackage[utf8]{inputenc} 
\usepackage[T1]{fontenc}    
\usepackage{hyperref}
\usepackage{ upgreek }
\usepackage{url}            
\usepackage{booktabs}       
\usepackage{amsfonts}       
\usepackage{nicefrac}       
\usepackage{microtype}      
\usepackage{lipsum}
\usepackage{float}
\usepackage{graphicx}
\usepackage{multirow}
\usepackage{array}
\usepackage{comment}
\usepackage{longtable}
\usepackage{ragged2e}
\usepackage{subfig}
\usepackage{amsmath}
\usepackage{longtable}
\usepackage{ amssymb }
\title{Active Learning on Medical Image}

\author{
  Angona Biswas   \\
  Research and Development Department, Pioneer Alpha,\\
Dhaka, Bangladesh\\
  \texttt{angonabiswas28@gmail.com} \\
   \And
    MD Abdullah Al Nasim   \\
  Research and Development Department, Pioneer Alpha,\\
Dhaka, Bangladesh\\
  \texttt{nasim.abdullah@ieee.org} \\
   \And
    Md Shahin Ali   \\
  Department of Biomedical Engineering, Islamic University, Kushtia-7003, Bangladesh\\
  \texttt{shahinbme.iu@gmail.com} \\
   \And
      Ismail Hossain \\
   University of Alabama at Birmingham\\
Alabama, USA\\
  \texttt{ihossain@uab.edu} \\
   \And
   Dr. Md Azim Ullah \\
   University of Memphis\\
  \texttt{mullah@memphis.edu} \\
  \And
Sajedul Talukder\\
University of Alabama at Birmingham\\
Alabama, USA\\
\texttt{stalukder@uab.edu}}

\begin{document}
\maketitle

\begin{abstract}

The development of medical science greatly depends on the increased utilization of machine learning algorithms. By incorporating machine learning, the medical imaging field can significantly improve in terms of the speed and accuracy of the diagnostic process. Computed tomography (CT), magnetic resonance imaging (MRI), X-ray imaging, ultrasound imaging, and positron emission tomography (PET) are the most commonly used types of imaging data in the diagnosis process, and machine learning can aid in detecting diseases at an early stage. However, training machine learning models with limited annotated medical image data poses a challenge. The majority of medical image datasets have limited data, which can impede the pattern-learning process of machine-learning algorithms. Additionally, the lack of labeled data is another critical issue for machine learning. In this context, active learning techniques can be employed to address the challenge of limited annotated medical image data. Active learning involves iteratively selecting the most informative samples from a large pool of unlabeled data for annotation by experts. By actively selecting the most relevant and informative samples, active learning reduces the reliance on large amounts of labeled data and maximizes the model's learning capacity with minimal human labeling effort. By incorporating active learning into the training process, medical imaging machine learning models can make more efficient use of the available labeled data, improving their accuracy and performance. This approach allows medical professionals to focus their efforts on annotating the most critical cases, while the machine learning model actively learns from these annotated samples to improve its diagnostic capabilities.


\end{abstract}

\section{Introduction}

Humans are dying of various types of diseases around the world every year. Some deadly diseases are excessively serious because if they are not detected at the early stage it becomes strenuous to save the patient's life. In that manner, the medical image screening process was invented and this scheme is vastly advanced now. Physicians were unable to see images of what was happening within a patient's body before November 8, 1895~\cite{bercovich2018medical}. As radiographs became a crucial component of medical diagnosis, new uses, and knowledge about disease representation soon accumulated, Roentgen was given the Nobel Prize in Physics in 1901 for X-ray. At the EMI research facilities in 1967, Sir Godfrey Hounsfield created the first CT scanner. On October 1, 1971, the first live patient was scanned.
The development of computed tomography (CT) involved creating a two-dimensional picture from radiography projections made from various angles. The first NMR pictures were published by Paul Lauterbur in 1973, and in 2003, he was awarded the 2003 Nobel Prize in the field of Physiology and Medicine. In contrast to CT, magnetic resonance imaging is based on distinct physical concepts. Then another ultrasound or US technology is mentionable. Ultrasound employs high-frequency sound waves that are sent into the body and are above the human hearing range. A positron emission tomography (PET) scan is an effective imaging procedure that can assist show how your tissues and organs' metabolisms or biochemical processes work~\cite{gallamini2014positron}. These are mostly used in the medical imaging process for the diagnosis of diseases. 

\begin{figure}
\centering
\includegraphics[height=6.2cm]{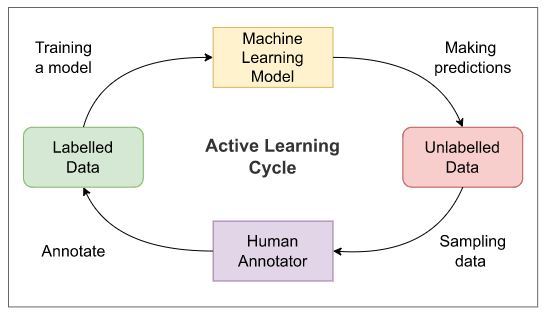}
\caption{The loop of training, testing, identifying uncertainty, annotating, and retraining is repeated until the model meets a predetermined performance level~\cite{erickson2017machine}.}
\label{ppr.JPG}
\end{figure}

Conventionally, a radiologist helps to diagnose or assist the doctor in predicting the deadly disease which is time-consuming and can be an unstable prediction. The involvement of technology is created a wide range of opportunities to diagnose medical images. Even more than a century ago, statistical methods of automated decision-making and modeling have been discovered (and updated) in a range of industries. The domain of Artificial Intelligence or Machine Learning created an automated process for the classification of deadly disease types and detection schemes which is a rapid and advanced procedure. Technology advances and federated learning make it possible for healthcare organizations to train machine learning models with private data without compromising patient confidentiality. We can get a concept of leveraging federated learning for x-ray images in the papers~\cite{talukder2022federated,puppala2022towards}. 

In~\cite{hossain2023collaborative} a collaborative federated learning system that enables deep-learning image analysis and classifying diabetic retinopathy without transferring patient data between healthcare organizations has been introduced. Along with image data, healthcare patients' statistical data can be used to train machine learning models in order to predict disease severity. In~\cite{talukder2022prediction,puppala2023machine} potential lead exposure at the zip code level is predicted  using machine learning on patients' Blood Lead Levels (BLL) dataset.

\begin{figure}
\centering
\includegraphics[height=6.5cm]{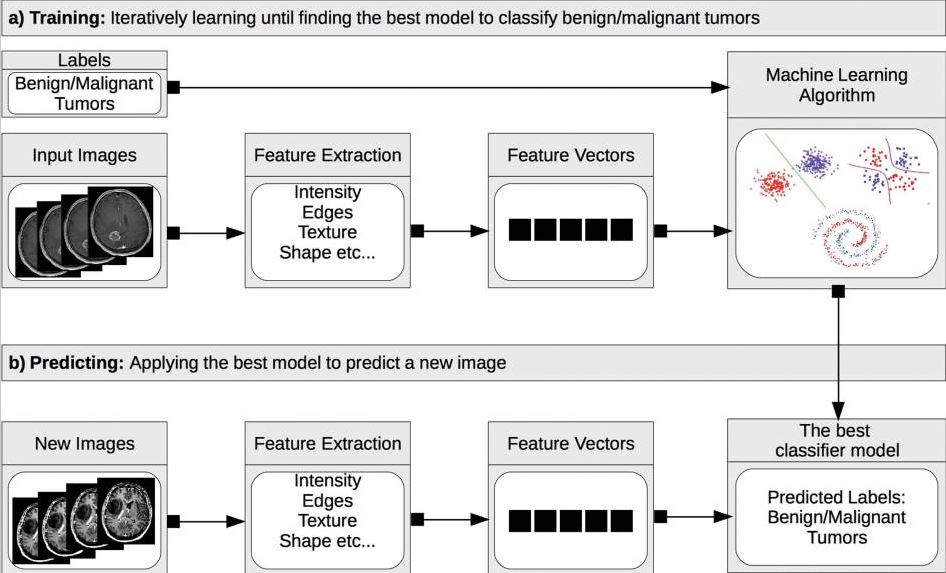}
\caption{The creation of machine learning models and their use in medical picture categorization problems~\cite{erickson2017machine}.}
\label{fig: kja-jpg}
\end{figure}

Figure~\ref{fig: kja-jpg} is depicting the process of how machine learning is utilized for medical image diagnosis. A machine learning algorithm learns from the training data and uses what it has learned to create a prediction if it is applied to a collection of data (in our case, tumor pictures) and information about these data (in our case, benign or malignant tumors)~\cite{erickson2017machine}. In past, various tremendous works are performed based on medical imaging. In~\cite{biswas2022mri}, a more precise automated brain tumor classification method using fuzzy C-means and ANN is suggested. MRI brain tumor is classified using ANN. This suggested method outperforms other current detection methods in terms of 99.8\% accuracy, 100\% sensitivity, and 99.59\% specificity. The main objective of this research ~\cite{makaju2018lung} is to evaluate several computer-aided methods, examine the most effective method already in use, pinpoint its drawbacks, and then propose a new model that improves on the most effective method already in use. The strategy employed involved grouping and listing lung cancer detection methods according to how well they could find the disease. Therefore, the goal of this research is to raise accuracy to 100\%. The lack of data produces poor results and raises additional unsolved questions regarding AI. It makes us realize that access is necessary for data preparation for machine learning and that even the greatest software would be meaningless without sufficient data filling. So, the problem arises when there are less amount of labeled medical images and the perfect detection or diagnosis is indispensable. This limitation is overcome by the process of ``Active Learning''. The conceptualization of active learning on medical images will be discussed in the following  section.

The annotators evaluate and identify the samples with a high level of uncertainty before sending them back. The model is then retrained using the freshly labeled data by ML developers before being tested once again on new unlabeled data. The early predictions that the model produces on the unlabelled data won't be accurate because the active learning pipeline starts with a short labelled sample.

This chapter will discuss the AI active learning (AL) method, which can assist human labelers by continuously arranging the unlabeled photos based on the level of information obtained, prompting the labeler to label the most informative image next.

\section{Literature Review}
Medical imaging continues to be very interesting in artificial intelligence (AI). The primary challenges to creating and utilizing AI algorithms in clinical settings involve the availability of suitably large, properly managed, and indicative training data that includes expert annotations or labeling. Willemink et al.~\cite{willemink2020preparing} outlines the basic procedures for readying medical imaging data for use in developing AI algorithms. They discuss the present shortcomings in data management and explore innovative methods for resolving the issue of limited data access. The task of classifying brain tumors is difficult, and Biswas et al.~\cite{biswas2022mri} introduced two clustering algorithms and an Artificial Neural Network (ANN) as techniques for classification. K-means and FCM are popular clustering methods used in medical image processing. 
A lot of fact-finding work is currently being done to improve radiology applications utilizing these algorithms to identify illness diagnosis system mistakes that might lead to very unclear medical therapies. The primary objective of Latif et al.~\cite{latif2019medical} is to emphasize the utilization of machine learning and deep learning methods in medical imaging. They aim to offer a summary of the current techniques employed in medical imaging for researchers and to underscore the benefits and drawbacks of these algorithms, while also considering potential future developments in the field. The use of artificial intelligence (AI) models is becoming more prominent in biomedical research and healthcare services. Castiglioni et al.~\cite{castiglioni2021ai} examine the challenges and issues that need to be addressed to effectively develop AI applications that function as clinical decision support systems in real-world situations.
In past, various improved experiment was done and successful research works were taken place for medical image diagnosis or classification. But there is a few research interlude. One of the major obstacles to the development and clinical use of AI systems is the absence of sufficiently large, vetted, representative data sets with expert labeling (eg, annotations). Before imagery may be used, several significant issues must be resolved. First, because picture annotation and tagging must be done, scalability is constrained. Second, each facility may need the replication and placement of significant computation resources. To resolve the less amount of labeled data, the concept of active learning is introduced which is acceptable for medical data. In contrast to other healthcare concerns, identifying COVID-19 requires AI-powered tools that implement active learning-based cross-population train/test models and incorporate multitudinal and multimodal data. Santosh et al.~\cite{santosh2020ai}'s main objective is to address this need.

\subsection {Machine Learning in the Context of Medical Images}

When there is a lot of information that has to be analyzed, diagnosing a brain tumor can be frustrating. The extraction of tumor areas from pictures becomes difficult due to the great degree of visual variety and similarities between brain tumors and normal tissues that characterize brain cancers~\cite{hossain2019brain,shah2019brain,hossainbrain}. The suggested study~\cite{pugalenthi2019evaluation} uses a Machine-Learning-Technique (MLT) to analyze the brain MRI slices and classify the tumor locations as low or high grade depending on the analysis. After a series of processes, the Gray Level Co-occurrence Matrix (GLCM) is then used to extract the essential information from the tumor area. The Support Vector Machine with Radial Basis Function (SVM-RBF) kernel is then used to create a two-class classifier, and its performance is checked against that of other classifiers like the Random-Forest and k-Nearest Neighbor. The proposed work's findings support that using a tool with the SVM-RBF to implement helps reach an accuracy of greater than 94\% on the benchmark BRATS2015 database. 
The main focus of this~\cite{biswas2022mri} research  is on accurately classifying brain tumors. In this chapter, two techniques for classifying MRI brain tumors are suggested by contrasting two key clustering algorithms with artificial neural networks (ANN). MRI images are initially pre-processed by a reseized and sharpening filter for algorithm 1. Second, k-means clustering is used to segment the photos. Thirdly, features are extracted using a 2D discrete wavelet transform and then compressed using principal component analysis. The final step is to use compressed features for ANN training, validation, and testing. For the second algorithm, C-means clustering is used along with a similar previous process. Finally, it is observed that the second algorithm is better than the first one. Segmentation of images has been addressed in \cite{karim2023c} which has numerous medical imaging applications.

An integrated AI and IoT technology that can aid in earlier breast cancer diagnosis. Mammograms are the primary method for finding breast cancer. In order to identify and treat breast cancer early and reduce the number of fatalities, several imaging techniques have been created. Additionally, numerous strategies for diagnosing breast cancer have been utilized to improve diagnostic precision~\cite{sivasangari2022breast}. The core principle of active learning stipulates that a machine learning approach may perform much better with even less training if provided the flexibility to choose the information it learns from~\cite{sinha2021prediction}. The percentage of persons afflicted by malignant and benign tumors has been predicted, and the findings have been shown using pictures. For an unbiased estimate in this suggested method, the authors have employed supervised learning techniques including K-Nearest-Neighbors, Naive Bayes, Decision Trees, and the Support Vector Machine Modelling Techniques. With an accuracy rate of 91.6\%, the findings showed that K-Nearest-Neighbors is the strongest predictor.

Among the increasing number of research on mass identification and segmentation, attention has recently been drawn to the automatic segmentation of breast ultrasound pictures into functional tissues. In the study of~\cite{xu2019medical}, we propose to employ convolutional neural networks (CNNs) to partition 3-dimensional breast ultrasound pictures into four primary tissues: skin, fibroglandular tissue, mass, and fatty tissue.

\begin{figure}
\centering
\includegraphics[width=10cm, height=9cm]{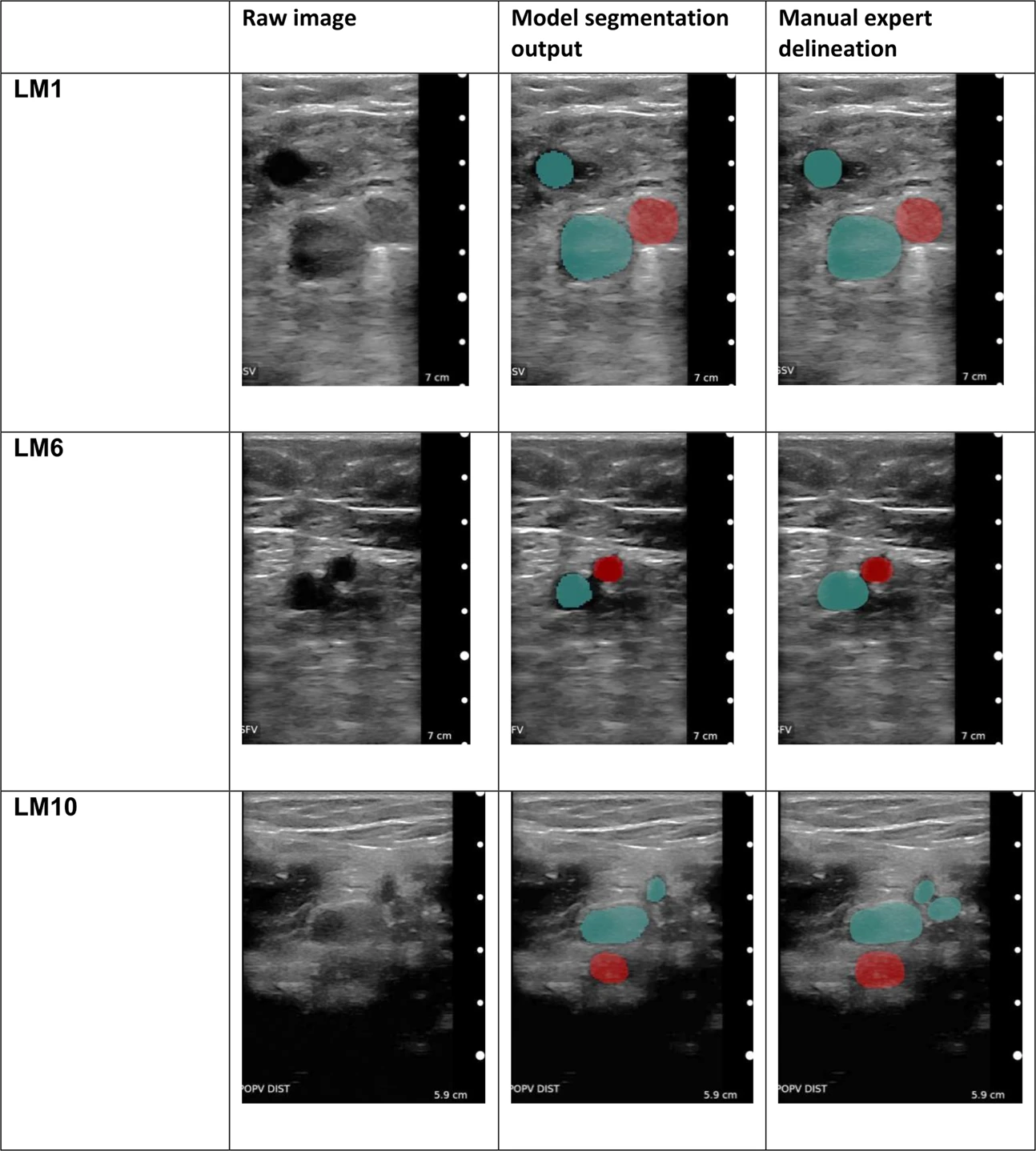}
\caption{Examples of high quality for the segmentation capabilities of our model~\cite{kainz2021non}}.
\label{fig:ultrasound.JPG}
\end{figure}

A blood clot called deep vein thrombosis (DVT), which is most frequently detected in the leg and can result in a catastrophic pulmonary embolism (PE). In a pre-clinical investigation of~\cite{kainz2021non} the authors gather photos and look into a deep learning method for automatically interpreting compressed ultrasound images. Our technique assists non-specialists in identifying DVT and offers direction for free-hand ultrasonography. We use ultrasound films from 255 volunteers to train a deep learning system, and we use 53 prospective NHS DVT diagnostic clinic patients and Thirty prospective German DVT hospital patients as our sample size for evaluation.Through all compressions, the segmentation is strong which is shown in~\ref{fig:ultrasound.JPG}. To rule out DVT, the venous region is assessed for total compressibility.

\subsection{Deep Learning in the Context of Medical Images}

\begin{figure}
\centering
\includegraphics[height=6.2cm]{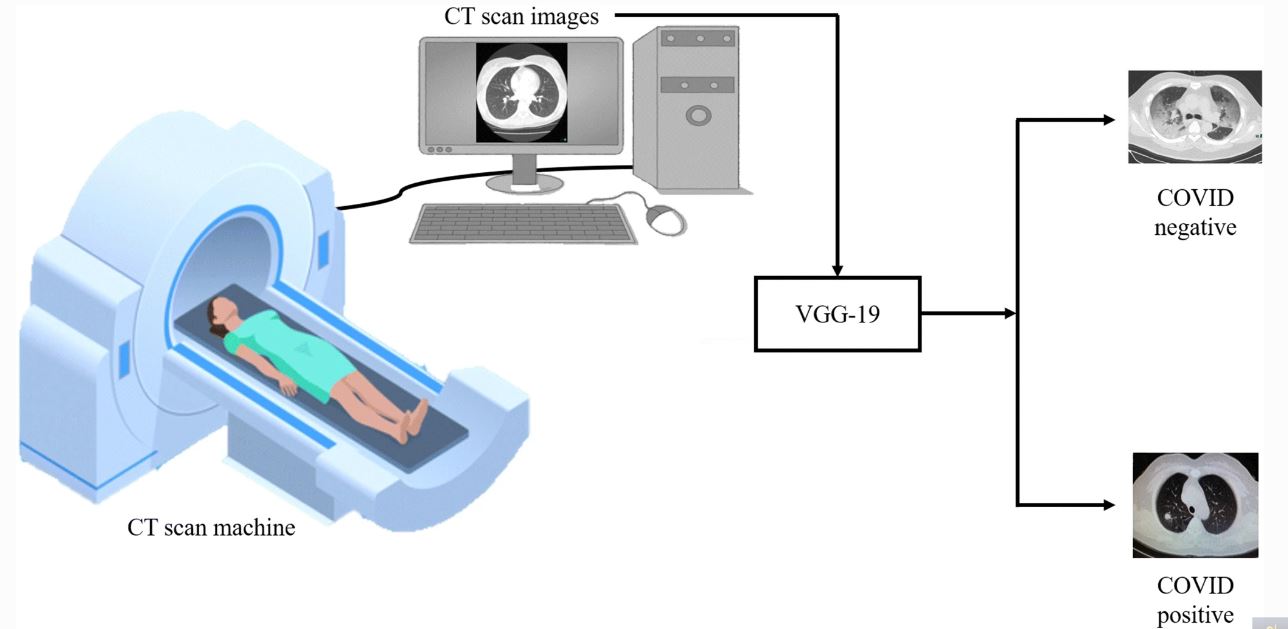}
\caption{Utilizing Deep Learning for automatic CT scan screening~\cite{shah2021diagnosis}}.
\label{fig: jgk.JPG}
\end{figure}

Pre-processing is employed in the~\cite{hasan2020classification} study to lessen the impact of intensity fluctuations across CT slices. The background of the CT lung image is then isolated using histogram thresholding. Each CT lung scan is subjected to feature extraction using a Q-deformed entropy technique and deep learning. Using a long short-term memory (LSTM) neural network classifier, the collected characteristics are categorized. The performance of the LSTM network is then considerably enhanced by merging all retrieved characteristics in order to accurately distinguish between COVID-19, pneumonia, and healthy patients. The obtained dataset, which includes 321 cases, may be classified with an accuracy of up to 99.68\%. The main goal of the publication~\cite{shah2021diagnosis} is to employ several deep learning approaches to distinguish between CT scan pictures of COVID-19 and non-COVID 19 patients. With an accuracy rate of 82.1\%, a self-developed model with the name CTnet-10 was created for the COVID-19 diagnostic. DenseNet-169, VGG-16, ResNet-50, InceptionV3, and VGG-19 are further models we examined. By having an accuracy of 94.52\%, the VGG-19 outperformed all other deep learning models. Figure~\ref{fig: jgk.JPG} shows that Deep Learning is effective for COVID-19 detection using lung CT Scan at the clinical stage. 

Artificial intelligence algorithms may be utilized to determine the existence and severity of the illness since there are obvious variations between the X-ray pictures of an infected individual and a healthy one. Pretrained models based on convolutional neural networks (CNN) have been widely utilized to classify Corona Virus Disease 2019 (COVID-19)~\cite{nasim2021prominence} using chest X-rays (CXR). This kind of successful work had been done by~\cite{shelke2021chest,goel2021optconet,jain2021deep} successfully during the pandemic situation.

This~\cite{tantawi2020bone} study suggests a two-stage categorization phase for identifying anomalies (fractures) in the upper extremity bones. For both classification phases, two convolutions neural network (CNN) models, ResNet-50 and Inception-v3, are examined. The outcomes show that the Inception model is better than all other classifiers for both phases. The average accuracy obtained surpasses that of previous investigations. The Mura dataset was used for all studies.

\begin{figure}
\centering
\includegraphics[height=4.2cm]{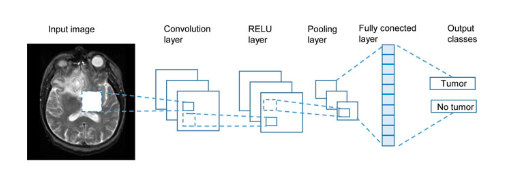}
\caption{Analysis of Medical Images Using Deep Learning (Brain tumor MRI diagnosis with the help of deep learning network)~\cite{ker2017deep}}.
\label{fig:ieeemedicalimage.png}
\end{figure}

Figure~\ref{fig:ieeemedicalimage.png} is showing how deep learning is effectively utilized for brain tumor detection. MR images are inputted into the deep learning network and the layers of the network extract the features finally the last layer provides the decision. 

\subsection{Issue of Inadequately Labeled Medical Data}

It is clear that the Artificial Intelligence domain needs data to learn the machine. After that, it can perform like a human or more perfectly. This is true for machine learning as well as deep learning. On the other hand, data annotation is necessary to ensure that systems produce reliable results and helps modules identify the components of speech and computer vision models. For some jobs, deep learning has proven to be more effective than traditional machine learning techniques, including those involving nature photographs. Deep learning has a variety of uses in medical imaging, such as the diagnosis of various diseases through the study of retinal fundus pictures and the categorization of lung and brain ailments. The capacity of deep learning to automatically extract the most pertinent characteristics for data interpretation and inference straight from the annotated data is what has led to its fast growth and deployment in such a wide range of applications. To train the deep learning model, however, a sizable amount of data with expert annotations is necessary. In reality, gathering a sizable volume of expert-annotated datasets in the field of medical images is difficult. The reason is that in order to appropriately identify the medical pictures, professionals such as radiologists or doctors who are occupied with clinical activities are needed in order to extract annotation from medical images. Additionally, medical pictures like CT, MRI, and PET are often difficult and expensive to collect, in contrast to natural image analysis, where the images might be simply taken with a regular camera. Deep learning is therefore increasingly important in medical picture applications to achieve generalizable learning from limited datasets.

\subsection{Active learning concept for medical images}
In a unique utilization of machine learning known as``active learning,'' fresh data points are labeled with the intended outputs by proactively querying a user (as well as another information source). It is occasionally referred to as the optimum design of experiments in the statistics literature. There are instances where there is a large amount of unlabeled data yet human labeling is expensive. Learning algorithms can actively ask the user or teacher for labels in this situation. Active learning is the name given to this recurrent supervised learning method.

The basic idea behind the active learner algorithm is that if an ML algorithm were given free rein to select the data it wishes to learn from, it could be able to achieve a greater degree of accuracy while utilizing fewer training labels. As a result, throughout the training phase, active learners are welcome to ask questions in an interactive manner. These requests are often sent as unlabeled data instances, and a human annotator is asked to label the instance. As one of the most effective instances of accomplishment in the living person paradigm, active learning is now included in that paradigm. The following are the three types of active learning:
1. Stream-based selective sampling
2. Pool-based sampling 
3. Membership query synthesis.
The two types of learning are more similar when they are active. Models are trained using both labeled and unlabeled data since it is a sort of semi-supervised learning. Semi-supervised learning is based on the hypothesis that labeling a limited sample of data may provide results that are as accurate as or perhaps more accurate than completely labeled training data. The only difficult part is figuring out what that sample is. During the training phase, active learning machine learning labels the data progressively and dynamically so that the algorithm may choose which label would be the most helpful for it to learn from.

\subsubsection{General Algorithm.}
\begin{enumerate}
    \item use the original training dataset to train the classifier
    \item determines the precision
    \item while (accuracy wanted accuracy):
    \item pick the most important data items (in general points close to the decision boundary)
    \item ask the human oracle for a label for the relevant data point(s).
    \item includes the aforementioned data point(s) in our first training dataset.
    \item Train the model again
    \item Update the accuracy calculation
\end{enumerate}

The technique of prioritizing the data that has to be labeled in order to have the most influence on training a supervised model is known as active learning. When there is too much data to label and smart labeling has to be prioritized because there is too much data to label, active learning can indeed be employed. In the case of medical images, there is less amount of labeled data that is generated regularly.

\section{Methodology}
\subsection{Case Description with Dataset information}
This section will discuss the details of the case which is based on MRI detection. When one brain tumor is diagnosed clinically, the radiologist evaluates the tumor's size, location, and effect on the surrounding area. It is obvious that accurate tumor-range diagnosis at the earliest stage will significantly increase the odds of survival for cancer patients. Brain tumors, traumatic brain injury, developmental abnormalities, multiple sclerosis, dementia, infection, stroke, and headache reasons can all be found with MRI technology \cite{al2022brain}. In order to use deep learning algorithms, a lot of photos must have excellent annotations. However, categorizing a lot of medical photos is difficult since it takes a lot of time and knowledge to annotate them. The lack of imaging studies and a lack of expert human annotations for images are the two key barriers to the usefulness of deep learning in diagnostic imaging. The research on active learning, on the other hand, suggests that it may be a viable strategy for developing a competitive classifier at a low annotation cost. With a reduced percentage of the training cohort, our proposed uncertainty sampling technique in this retrospective study utilizes transfer as well as active learning to provide consistent test results.

\begin{figure}
\centering
\includegraphics[height=4.8cm]{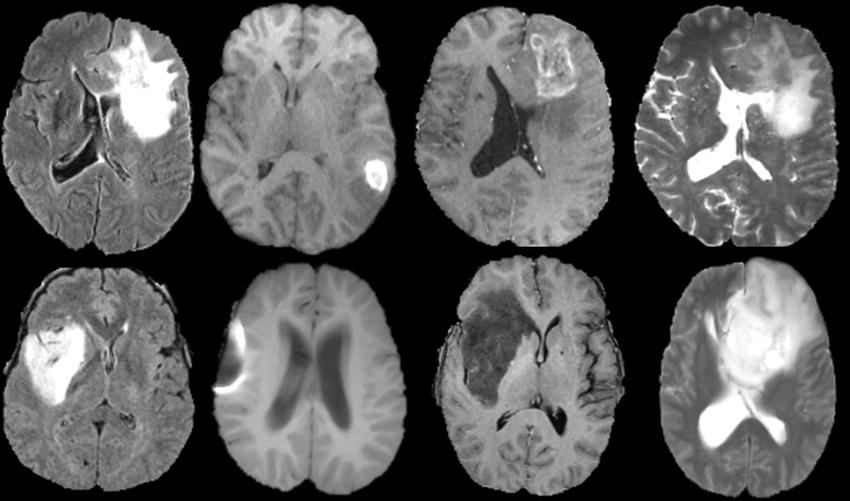}
\caption{BRATS Dataset~\cite{menze2014multimodal}}.
\label{fig: brats.jpg}
\end{figure}

However, the size of the dataset with annotations has a significant impact on the performance of deep learning approaches. Given the complexity and abundance of medical data, it is very difficult to label a significant number of medical pictures. In this study, the authors present an innovative transfer learning-based active learning framework to lower the annotation cost while retaining the robustness and stability of the model performance for classifying brain tumors. In this retrospective study, the authors utilized the magnetic resonance imaging (MRI) training dataset of 203 patients and the validation dataset of 66 patients as the baseline to train and fine-tune our model. The BRATS 2019 dataset~\cite{bakas2017advancing,menze2014multimodal} which contains 335 individuals with brain tumor diagnoses, served as the basis for this study (259 patients with HGG and 76 patients with LGG). The median age of the 240 patients with age data available is 60.31 years. The dataset is shown in Figure~\ref{fig: brats.jpg}. Four MRI sequences—T1-weighted, post-contrast-enhanced T1-weighted (T1C), T2-weighted (T2), and T2 fluid-attenuated inversion recovery (FLAIR) volumes—are included in each patient's MRI scan collection. The dataset underwent skull-striping as preprocessing, was interpolated to a uniform isotropic resolution of 1 mm3, and registered to SRI24 space with a dimension of 240 240 155. Four labels are included in the dataset's annotations: background, gadolinium-enhancing tumor, peri-tumoral edema, and the center of the necrotic and non-enhancing tumor. The whole tumor region is represented by the area denoted by the final three of the four labels. In order to use the suggested methodology in this study~\cite{hao2021transfer,tonmoy2019brain}, we randomly selected 20 slices from each patient's axial plane MRI scan that included the tumor site while maintaining the T1, T1C, and T2 channels for each slice. We selected the T1, T1C, and T2 channels from a total of four channels based on the outcomes of the initial trials. The pre-trained AlexNet requires three-channel input. The 6,700 2D 3-channel slice dataset was further divided into a test set (203 patients), a validation set (66 patients), and a training set (66 patients). The ratio of HGG patients to LGG patients in each of the three cohorts is the same as it is for the entire dataset. Slices with LGG tumors were all labeled as 0, whereas slices with HGG tumors were all labeled as 1. To suit the pre-trained CNN, the pictures were scaled down from 240*240 pixels to 224*224 pixels.

\subsection{MRI Pre-processing} Pre-processing MR images is a critical step in any workflow for quantitative analysis. These preprocessing steps might be made up of many procedures, each of which aims to either enhance the quality of the image or standardize its geometrical and intensity patterns~\cite{manjon2017mri}. Preprocessing steps that were used in an experiment will be mentioned in this paragraph.

The data were registered to SRI24 space with a dimension of 240 240 155, interpolated to a uniform isotropic resolution of 1 mm3, and preprocessed using skull-striping. The dataset has four labels as part of its annotations: background, gadolinium-enhancing tumor, peri-tumoral edema, and necrotic and non-enhancing tumor core. The entire tumor region is indicated by the final three labels on the four labels.

\subsection{Framework for Active Learning Based on Transfer Learning knowledge}
A CNN must be trained from scratch (with random initialization), which takes a lot more time and computing power than using a CNN that has already been pre-trained on a sizable dataset~\cite{gu2018recent, li2021survey, yamashita2018convolutional}. Finetuning and freezing are the two primary transfer learning possibilities. In fine-tuning, the weights and biases of a pre-trained CNN are adopted in place of random initialization, and a traditional training procedure is then carried out on the target dataset. We think of the pre-trained model CNN layers as a static feature extractor in the freezing case. In this case, we allow the fully connected layers to be modified across the target dataset while authors freeze the weights and biases of the convolutional layers we want to use. 
The convolutional layers do not have to be the only frozen layers. Any subset of convolutional or fully connected layers can be chosen as the frozen layers; however, it is customary to only freeze the shallower convolutional levels. In this study, the ImageNet Large-Scale Visual Recognition Challenge (ILSVRC) dataset~\cite{russakovsky2015imagenet}, which contains real-world pictures, is used to pre-train the CNNs.The authors choose fine-tuning as our transfer learning approach since the ImageNet dataset and our target medical picture domain are very different from one another.

\begin{figure}
\centering
\includegraphics[height= 6.00cm]{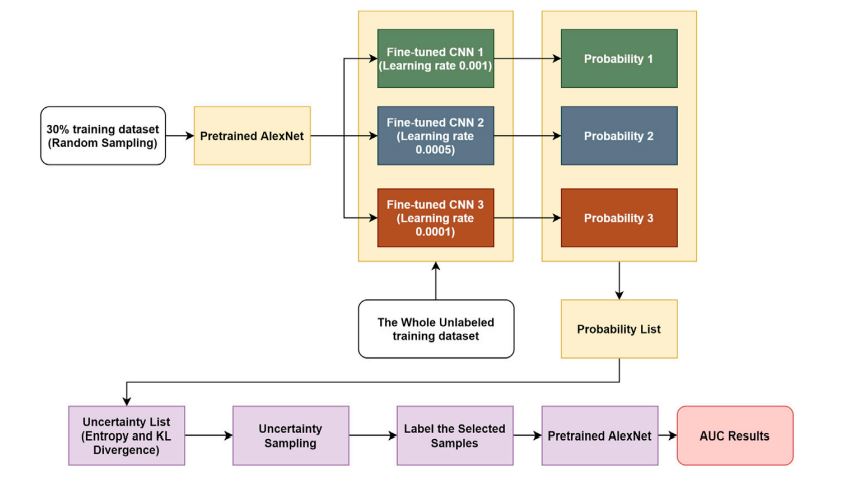}
\caption{Workflow of the suggested active learning framework based on transfer learning~\cite{hao2021transfer}}.
\label{fig: active.JPG}
\end{figure}

The authors chose the pre-trained AlexNet and modified it on the BRATS 19 dataset to lower the annotation cost. Three max-pooling layers, three fully connected layers, and five convolutional layers make up AlexNet. The depth of AlexNet is suitable for classifying brain tumors and is significantly shallower than other benchmark CNNs (such as ResNet and VGG), which promotes faster convergence and requires fewer processing resources.

In this study, the authors provide a novel transfer learning-based active learning framework to lower the annotation cost while retaining the stability and robustness of CNN efficacy for classifying brain tumors. The procedure for our active learning is shown in Figure~\ref{fig: active.JPG}.
Authors assume that there are labeled and unlabeled subgroups in the training dataset. Finding the most insightful examples throughout the whole training set is the objective; these samples may or may not coincide with the labeled training subset. There are four steps in the workflow: 1) Authors chose 30\% of the training data at random for the labeled training set, assuming the other 70\% of the samples were unlabeled. The pre-trained AlexNet was then fine-tuned using the 30\% labeled training subset, and the learning rate was adjusted to different levels (i.e., 0.001, 0.0005, and 0.0001). This phase allowed us to get three improved CNNs. 2) To determine the classification probabilities of each sample in the complete training dataset, we employed these fine-tuned CNNs.There is no need for labels in this phase because the CNNs solely use forward propagation to calculate outputs. 3) After each training sample generated three predicted possibilities in step 2, the calculation is done by pairwise KL divergence and individual entropy. The entropy and KL divergence of each sample is added to create the uncertainty score. An uncertainty score list for the full training dataset was achieved using this method. 4) Authors picked 30\% of the training cohort, which was made up of the most informative data, and sorted the uncertainty score list in descending order. This chosen subset needed to be labeled, therefore it was utilized to improve a pre-trained AlexNet.The maximum training size required is 30 + 30= 60\% (40\% reduction in training size) of the whole training cohort if there was no overlap between the initially labeled training subset (30\%) and the most informative subset identified (30\%). The maximum training size required is just 30\% (70\% reduction in training size) of the whole training cohort if the most informative samples identified are precisely identical to the original labeled training subset (30\%). In other words, this suggested transfer learning-based active learning architecture can save between 40 and 70\% of annotation costs (on average, 55\%).

\subsection{Case Report and Analysis}

On a different test dataset of 66 patients, using this suggested strategy, the model produced an area under the receiver operating characteristic (ROC) curve (AUC) of 82.89\%, which was 2.92\% higher than the baseline AUC while saving at least 40\% of the labeling cost. The authors built a balanced dataset and ran the same process on it to further investigate the robustness of our strategy. The model's AUC of 82\%, compared to the baseline's AUC of 78.48\%, confirms the stability and robustness of this suggested transfer learning enhanced with an active learning framework while drastically decreasing the amount of training data.

\begin{figure}
\centering
\includegraphics[width=13cm, height= 6.00cm]{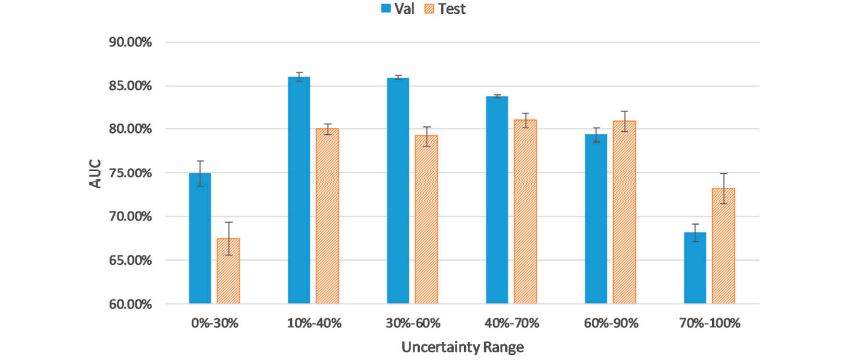}
\caption{Performance of CNN on samples from various uncertainty intervals~\cite{hao2021transfer}}.
\label{fig: train.JPG}
\end{figure}

The authors of Results of Using Transfer Learning show that transfer learning is a successful strategy and enhances our base models. The top 10\% of certain and unsure cases are shown in AUC Results of selecting a different range of uncertainty distribution to be uninformative, and hence leaving them out allows the models to generalize more effectively. We will conduct experiments to demonstrate how our uncertainty sampling strategy enhances the baseline with a sample size fixed at 30\% in AUC Results of the Uncertainty Sampling Method. Finally, we will illustrate the following in the AUC results of the Uncertainty Sampling Method on a balanced dataset: 1) Our sampling technique works whether the dataset is balanced or unbalanced. 2) Our sampling strategy's improvement of the baseline is neither accidental nor the product of chance.

The amount of labeled training data needed may be drastically decreased while retaining a high level of tumor classification accuracy using a transfer learning-based active learning architecture. 
As shown in Figure~\ref{fig: train.JPG} the top 30\% of definite instances or the top 30\% of uncertain examples are chosen to lower the AUC findings for the validation (and test) group. We removed the top 10\% (most uncertainty scores) and bottom 10\% (lowest uncertainty total score) samples to find outliers only with the lowest training values. According to Fig. reftrain.JPG, the uncertainty range / -40\% boosts AUC values by 12.51\% when compared to the band of 0-30\%. Similar to the interval of 70-100 percent, the ambiguity range of 60-90 percent increases AUC by 7.72 percent.

\section{Conclusion}
The goal of medical imaging is to identify diseases and gain insights into the functioning of organs and anatomy. Various medical procedures, such as identifying lung tumors, spinal deformities, and artery stenosis detection, can be carried out through medical imaging. Image processing techniques such as enhancement, segmentation, and denoising, as well as machine learning methods, can improve the performance of medical imaging. Machine learning algorithms can classify or measure image properties optimally if given enough labeled data, which is a challenge for medical images due to a lack of labeled data. To address this challenge, the chapter discusses the active learning concept, which involves using machine learning to train on relevant data and improve accuracy for quick medical diagnosis, specifically for medical MR images.


\bibliographystyle{unsrt}
\bibliography{main.bib}

\end{document}